\documentclass[pre, twocolumn,showpacs,preprintnumbers,amsmath,amssymb]{revtex4}
 \usepackage{graphicx}
 \usepackage{amssymb}
 
\newcommand{\vect}[1]{\mathbf{#1}}

\begin{document}	
\title{Numerical calculations of a high brilliance synchrotron source and on issues with characterizing strong radiation damping effects in non-linear Thomson/Compton backscattering experiments}
\author{A. G. R. Thomas$^{1,2}$, C. P. Ridgers$^3$, S. S. Bulanov$^{4}$,  B. J. Griffin$^{2}$, S. P. D. Mangles$^5$}
\address{$^1$Centre for Ultrafast Optical Science, University of Michigan, Ann Arbor, MI 48109,  US.}
\address{$^2$Department of Nuclear Engineering and Radiological Sciences, University of Michigan, Ann Arbor, MI 48109, USA.}
\address{$^3$Clarendon Laboratory, University of Oxford, Parks Road, Oxford OX1 3PU, UK.}
\address{$^4$University California Berkeley, Berkeley, CA 94720 USA.}
\address{$^5$Blackett Laboratory, Imperial College London, London SW7 2AZ, UK}
\date{\today}
\begin{abstract}
A number of theoretical calculations have studied the effect of radiation reaction forces on radiation distributions in strong field counter-propagating electron beam-laser interactions, but could these effects -- including quantum corrections -- be observed in interactions with realistic bunches and focusing fields, as is hoped in a number of soon to be proposed experiments? We present numerical calculations of the angularly resolved radiation spectrum from an electron bunch with parameters similar to those produced in laser wakefield acceleration experiments, interacting with an intense, ultrashort laser pulse. For our parameters, the effects of radiation damping on the angular distribution and energy distribution of \emph{photons} is not easily discernible  for a ``realistic'' moderate emittance electron beam. However, experiments using such a counter-propagating beam-laser geometry should  be able to measure such effects using current laser systems through measurement of the \emph{electron beam} properties. In addition, the brilliance of this source is very high, with peak spectral brilliance exceeding $10^{29}$ photons$\,$s$^{-1}$mm$^{-2}$mrad$^{-2}(0.1$\% bandwidth$)^{-1}$ with approximately 2\% efficiency and with a peak energy of 10 MeV.
\end{abstract}
\maketitle
\section{Introduction}
The recent development of ultra-high intensity laser systems has generated a great amount of interest in a class of well known theoretical problems involving the interaction of strong fields with relativistic electron beams that have not been experimentally demonstrated. Relativistic electron beams are regularly measured in experiments by laser wakefield acceleration \cite{Tajima_PRL_1979,mangles,geddes,faure} and are characterized by being of relatively high current density in short bunches. In laser wakefield acceleration, oscillations of the electrons in the electromagnetic fields of electron plasma cavities created by laser driven ponderomotive expulsion have been shown to result in extremely bright sources of x-rays \cite{Rousse_PRL_2004,ISI:000242538700029,ISI:000254024500038,ISI:000253724500017,Thomas_POP_2009,Kneip_NP_2010}.

\begin{figure}[htbp]
\begin{center}
\includegraphics[width=0.5\textwidth]{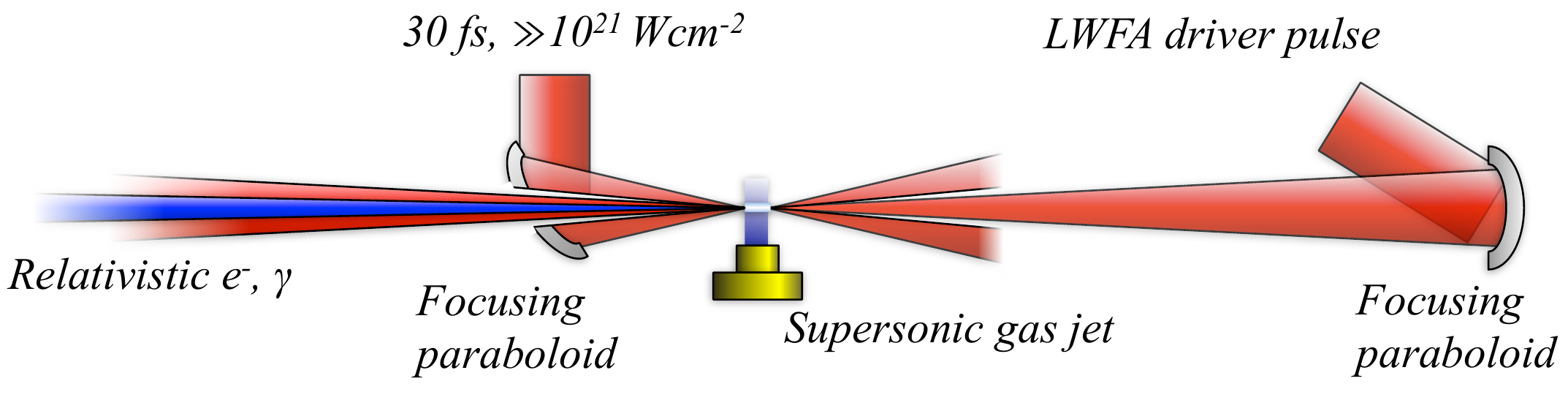}
\caption{Schematic of counter-propagating laser-beam interaction geometry using laser wakefield accelerated electrons.}
\label{figureminus1}
\end{center}
\end{figure}

Another proposed source of radiation using the wakefield accelerated electron beam is Thomson or Compton backscattering from a second laser \cite{Esarey_PRE_1993,Hartemann_PRL_1996,Salamin_JPA_1998,Ueshima_LPB_1999,Avetissian_PRE_2002,Lee_OE_2003,Brown_PRSTAB_2004,Hartemann_PRE_2005,Koga_POP_2005,Albert_POP_2011}. In this scheme a counter-propagating laser is used as a short wavelength undulator for producing high brightness, monochromatic gamma rays. An undulator in a conventional synchrotron is characterized by a strength parameter $K$ that characterizes the oscillation amplitude relative to wavelength. For small $K$, the radiation is monochromatic. For large $K$ the radiation is characterized by a synchrotron-like spectrum \cite{Jacksonbook}. In the counter-propagating laser scheme, the field strength parameter (normalized peak vector potential) $a_0=|eF_0|/m_ec\omega_0$ is analogous to $K$. $F_0$ is the peak electric field strength of a laser with central angular frequency $\omega_0$. For a laser with $a_0\ll1$ ($I\lambda^2\ll 10^{18}$~Wcm$^{-2}$), the radiation is monochromatic. For $a_0>1$ harmonics in the radiation spectrum start to appear, and for $a_0\gg1$ the spectrum becomes broad. For linear polarization of the laser there is also longitudinal motion due to the Lorentz force, and therefore downshifting of the fundamental frequency occurs \cite{He_PRL_2003,ISI:000182450200081}. The monochromatic regime using a laser wakefield accelerated electron bunch has been proposed as a good source for applications \cite{He_PRL_2003,Hartemann_PRSTAB_2007,Albert_PRSTAB_2010,Albert_PRSTAB_2011}. In addition, experiments using this counter propagating geometry with a very high intensity laser (Fig \ref{figureminus1}) should be an interesting testbed for studying radiation reaction forces and non-linear quantum electrodynamics \cite{Bulanov_NIMPR_2011}, due to the high field strength in the  electron rest frame.

The transverse component of the laser vector potential is Lorentz invariant,  so the radiation emission of an $a_0\leq 1$ interaction  is very different to an $a_0\gg1$ interaction independent of the reference frame (and therefore electron energy, in the colliding geometry). The emission of photons in such processes clearly indicates that a force should be applied to the electron to conserve momentum. Conversely, the electric field strength is not a Lorentz invariant, and hence the electron energy in this geometry may be crucial to determining whether the field is quantum electrodynamically strong or not.

In this paper, radiation damping effects on the full angular and energy distribution of photons produced in the counter-propagating geometry interaction between a tightly focused $\sim10^{22}$~Wcm$^{-2}$  ultrashort pulse with a electron beam is studied by solving modified classical equations of motion numerically and generating spectra with a numerical radiation spectrometer \cite{Thomas_PRSTAB_2010}. 
 The layout of the manuscript is as follows: First we parameterize the interplay between the field strength $a_0$ and electron energy $\gamma m_ec^2$ in the colliding pulse geometry, and identify the regime relevant to near term experiments where  radiation damping is strong but quantum electrodynamic effects are relatively small. Next we introduce the numerical model for calculating both the electron dynamics and radiation spectra. We then proceed to calculate the $\gamma$-ray spectrum with ``realistic'' conditions, then examine the effect of radiation reaction on the photon and electron phase-spaces. Finally, we show that semi-classical corrections to the radiation reaction force may be observable in experiments.
\section{Parameterizing strong field interactions}
\subsection{Radiation reaction force effects}
Although properly described by quantum electrodynamics, the radiation force has a classical form that is self-consistent within the limits that the acceleration timescale is much larger than $\tau_0=2e^2/3mc^3=6.4\times10^{-24}$~s \cite{Dirac_PRSA_1938,Rohrlich_PRE_2008}. It is principally a damping of motion due to loss of momentum to the radiation. The Lorentz-Abraham-Dirac equation is a third order differential equation of motion for a charged particle in the presence of accelerating forces, and includes the change of momentum due to the radiation generated by the charge. The force on an electron is given in covariant form by:
 \begin{equation}
\frac{d}{d\tau}v^{\mu}=-\frac{e}{m_e}\left(F^{\mu\nu}v_{\nu}+\tau_0D^{\mu}\right)\;,
\label{LAD}
\end{equation}
where $D^{\mu}$ is the radiation reaction (damping) force, $F^{\mu\nu}=\partial^\mu A^\nu-\partial^\nu A^\mu$ is the electromagnetic field tensor and $v_\nu = {dx^\nu}/{d\tau}=\{\gamma c, -\gamma \mathbf{v}\}$ is the particle four-velocity. 

The radiation reaction force, according to the Lorentz-Abraham-Dirac model, is a source of much controversy precisely because it is a third order differential equation, which allows for self-accelerating solutions that do not conserve energy, for example. Various authors have reformulated the equation to eliminate the third order term (See Sokolov \cite{Sokolov_POP_2009},  Hammond \cite{Hammond_PRA_2010} and references within). These are generally identical to first order in $\tau_0$ (and are therefore  basically all equivalent to the Landau-Lifshitz  form of the radiation reaction force \cite{Landau_Lifshitz}), but are otherwise not identical. The modified force can be written in the form \cite{Rohrlich_PRE_2008}:
 \begin{equation}
\frac{d}{d\tau}v^{\mu}=-\frac{e}{m_e}\left[F^{\mu\nu}v_{\nu}
-\tau_0P^{\mu\alpha}\frac{d}{d\tau}\left(F_{\alpha}^{\nu}v_{\nu}\right)\right]\;.
\label{PLAD}
\end{equation}
where $P^{\mu\alpha}=\eta^{\mu\alpha}+{v^{\mu}v^{\alpha}}/{c^2}$ and $\eta^{\mu\nu}$ is the Minkowski metric tensor with trace -2. In Ref. \cite{Bulanov_PRE_2011}, several examples were given, which show that the solutions of the Lorentz-Abraham-Dirac model and equation \ref{PLAD} are identical in the classical regime. 

One of the interesting phenomena arising from this laser-electron interaction is that {the radiation damping is theoretically predicted to be so extreme that for a sufficiently intense laser, the electron beam may lose almost all its energy in the interaction time}. In particular, Koga et al. studied the effect of radiation damping on the radiation spectrum \cite{Koga_POP_2005}. Di Piazza et al. also studied the effect of radiation damping on the angular distribution of radiation \cite{Piazza_PRL_2009}. The effects of `real world' conditions on the radiation spectrum emitted have also been somewhat previously studied, for example the effects of higher-order field corrections for tightly focused pulses \cite{Hartemann_AJS_2000,Lee_EPL_2010}.

Radiation damping can be parameterized by considering the energy loss of the electron due to the most significant damping term \cite{Koga_POP_2005,Bulanov_NIMPR_2011}. Here we proceed from equation \ref{PLAD}, where ignoring terms of $\tau_0^2$ and higher and the Schott term, the damping contribution can be written in the form \cite{Rohrlich_PRE_2008}:
 \begin{equation}
\frac{d}{d\tau}v^{\mu}=-\frac{e}{m_e}F_\alpha^{\nu}v_{\nu}\left[\eta^{\alpha\mu}-\tau_0\frac{e}{m_ec^2}v_{\nu}v^\mu F^{\alpha\nu}\right]\;,
\label{PLADdamped2}
\end{equation}

The electromagnetic four-force can be written in the form:
\begin{eqnarray}
F^{\alpha\nu}v_{\nu} = -\frac{dA^\alpha}{d\tau}+v_\nu\partial^\alpha A^\nu\;.
\end{eqnarray}

For the case of a linearly polarized plane wave, $A^\mu = \Re \left[\{A_0\}^\mu e^{i\kappa_\alpha x^\alpha}f(\kappa_{\beta} x^\beta/\omega_0t_L)\right]$ where $\kappa_\alpha$ is the four-wave-vector $\kappa_\alpha=\omega_0 \{ 1,-\vect{\hat{k}}/c\}$, $f(\kappa_{\alpha} x^\alpha/\omega_0t_L)$ is a function describing the temporal envelope and $t_L$ is the pulse duration, interacting with a counter-propagating  electron with initial Lorentz factor $\gamma_0$ obeying $a_0\ll\gamma_0\ll (a_0\omega_0\tau_0)^{-1/2}$, the zeroth component is well approximated by:
\begin{equation}
\frac{d\gamma}{d\tau} = -\gamma \tau_0\frac{d a^\mu}{d\tau}\frac{d a_\mu}{d\tau}\;,
\label{simple_damping}
\end{equation}
where $a^\mu=eA^\mu/m_ec$. The condition on $\gamma$ is so that the longitudinal Lorentz force is minimized but radiation damping does not affect the transverse oscillations of the electron. For a slowly varying gaussian envelope, i.e. $(1/f)df/d\tau\ll\kappa_\mu v^\mu$ 
with $f = \exp(- (\kappa_{\alpha} x^\alpha/\omega_0t_L)^2)$, and averaging over the fast oscillations, we can integrate to obtain the 
total energy loss by the particle: 
\begin{equation}
\frac{\Delta\gamma_\infty}{\gamma_0} = \frac{\sqrt{\frac{\pi}{2}}\tau_0t_L\omega_0^2\gamma_0a_0^2}{1+\sqrt{\frac{\pi}{2}}\tau_0t_L\omega_0^2\gamma_0a_0^2}\;,
\end{equation}

This is similar to the result in Ref. \cite{Bulanov_PRE_2011}, but with a different definition for the pulse duration because here $t_L$ is close to the full-width-at-half-maximum duration commonly used in experiments. From this expression, we can define a parameter $\psi=10\sqrt{\frac{\pi}{2}}\tau_0t_L\omega_0^2\gamma_0a_0^2 t_{rad}$, for a particular characteristic timescale for radiation damping $t_{rad}$, such that:
\begin{equation}
\frac{\Delta\gamma_\infty}{\gamma_0} = \frac{0.1\psi (t/t_{rad})}{1+0.1\psi (t/t_{rad})}\;.
\end{equation}
which clearly defines strong radiation damping for $\psi\geq1$ and weak radiation damping for $\psi\ll1$. Here we choose $t_{rad}=2\pi/\omega_0$ -- that is to say a laser period -- which is slightly different from the choice of Koga et al. \cite{Koga_POP_2005}, who chose the pulse duration for $t_{rad}$. However, we have also added a factor of 10 into $\psi$, which is such that $\psi=1$ corresponds to a 10\% energy loss in a single cycle, which therefore results in a condition similar to that of Koga et al. \cite{Koga_POP_2005}, since they considered an approximately 10 cycle pulse. In addition, a 10\% loss in a single cycle can reasonably be defined as the threshold of ``significant'' damping. Hence:
\begin{equation}
\psi= 10\sqrt{2\pi^3}\omega_0\tau_0\gamma_0a_0^2\;.
\end{equation}
For a 800 nm laser, $\psi=1.2\times10^{-6}\gamma_0a_0^2$. 
The condition $\psi=1$ leads to the condition for the laser pulse vector potential, \textit{i.e.} the strong radiation damping regime is realized for
\begin{equation}\label{a_rad}
a_0>a_{rad}=\left(10\sqrt{2\pi^3}\omega_0\tau_0\gamma_0\right)^{-1/2}.
\end{equation}
\subsection{Quantum electrodynamics effects}
Quantum electrodynamically strong interactions are 
parameterized by a relativistically and gauge invariant parameters  $\chi_e= ||F_{\mu\nu}v^{\nu}||/(c E_{cr})$ and $\chi_\gamma= ||F_{\mu\nu}\hbar k^{\nu}||/(m_e c E_{cr})$ \cite{Ritus_1979}, where $\hbar k^\nu$ is the four-momentum of  a photon and $E_{cr}= m_e^2c^3/e\hbar=1.32\times 10^{18}$ Vm$^{-1}$ is the Schwinger or critical field of quantum electrodynamics. These parameters determine the rates of photon creation by an electron  or an electron-positron pair creation by high-energy photon in a strong electromagnetic field, the latter being the Breit-Wheeler process 
\cite{Breit_PR_1934}. The photon emission probability for $\chi_e\ll 1$ is $\approx \left(5\alpha m_e^2/2\sqrt{3}p_0\right)\chi_e$ and for $\chi_e\gg 1$ is $\approx \left[14\Gamma(2/3)\alpha m_e^2/27p_0\right]\left(3\chi_e\right)^{2/3}$, where $p_0$ is the electron energy and $\Gamma(z)=\int_0^\infty t^{z-1}e^{-t}dt$ is the Euler gamma function, and $\alpha=e^2/4\pi\epsilon_0\hbar c=1/137$ is the fine structure constant \cite{Ritus_1979}. The pair production probability by a photon for $\chi_\gamma\ll 1$ is $\approx \left(3\sqrt{3}\alpha m_e^2/16 \sqrt{2} k_0\right)\chi_\gamma\exp\left(-8/3\chi_\gamma\right)$ and for $\chi_\gamma\gg 1$ is $\approx \left[15\Gamma^4(2/3)\alpha m_e^2/28\pi k_0\right]\left(3\chi_\gamma\right)^{2/3}$, where $k_0$ is the photon energy \cite{Ritus_1979}. Previously it was shown that    extremely high intensity counter propagating laser pulses could lead to prolific pair production  \cite{Bell_PRL_2008,Fedotov_PRL_2010,Bulanov_PRL_2010}.

 For multi-100 TW lasers, such as the {\sc  Hercules}  \cite{Yanovsky:2008} or Astra Gemini \cite{Hooker_JPIV_2006} lasers, with focused field strength $|E|\sim 10^{-3}E_{cr}$, interaction with GeV energy electron beams should be sufficient to achieve $\chi_e\sim 1$ \cite{Schwinger_PR_1951,Sokolov_PRL_2010,Sokolov_PRE_2010}. However the conversion of emitted photons into electron-positron pairs will be suppressed due to the $\exp\left(-8/3\chi_\gamma\right)$ in the expression for the probability for $\chi_\gamma\ll 1$. 

A notable experiment in a similar geometry, using the 46~GeV electron beam from the the Stanford Linear Accelerator (SLAC) colliding with a laser with intensity of $I_0\sim 10^{18}$~Wcm$^{-2}$,  was an important demonstration of non-linear quantum electrodynamics (multi photon Breit-Wheeler pair production)
\cite{Burke_PRL_1997}.  A simplified version of the parameter $\chi_e$ for the situation of an electron beam with energy $E=\gamma_0m_ec^2$ colliding with a laser field with field strength parameter $a_0$ can be written as \cite{Sokolov_PRE_2010}
\begin{equation}
\chi_e = \frac{2\hbar}{m_ec^2}\omega_0\gamma_0 a_0\;.
\end{equation}
For an 800~nm laser system, this gives $\chi_e = 6\times10^{-6} \gamma_0a_0$. For the SLAC experiment (using a 527 nm laser), the small $a_0$ ($a_0<1$) was compensated by the high beam energy ($\gamma_0\sim10^5$), so that $\chi_e \approx 0.4$. 

 Quantum electrodynamics effects may be considered significant when the energy of the emitted photons becomes of the order of the electron energy, $\hbar\omega\gtrsim \gamma_0 m_ec^2$. For a head-on collision of an electron and a laser pulse, a characteristic emitted photon energy is $\hbar\omega\approx \hbar\omega_0 a_0 \gamma_0^2$ \cite{Bulanov_NIMPR_2011}, which corresponds to the condition $\chi_e\sim1$. Hence,  quantum electrodynamics effects may be considered to be strong for a field strength of:
\begin{equation} \label{a_Q}
a_0>a_Q=\frac{m_e c^2}{2\hbar\omega_0\gamma_0}.
\end{equation} 

However, quantum effects in the radiation damping  of electrons becomes noticeable for much lower laser field strengths. It is well known
 \cite{Ritus_1979,Erber_RMP_1966,Kirk_PPCF_2009,Sokolov_PRE_2010}, that the classical description of an electron radiating in a strong electromagnetic field overestimates the total emitted power. It is connected with the fact that in the quantum description the emitted photon energy may not exceed the electron energy, whereas the classical approach does not have such a restriction. This effect can  be approximately taken into account by introducing a function $g(\chi_e)$ into the expression for the total power of emitted
radiation \cite{Erber_RMP_1966,Kirk_PPCF_2009,Sokolov_PRE_2010}. $g(\chi_e)$  enters  the equation of motion by modifying the expression for the radiation reaction force as:
 \begin{equation}
\frac{d}{d\tau}v^{\mu}=-\frac{e}{m_e}F_\alpha^{\nu}v_{\nu}\left[\eta^{\alpha\mu}-g(\chi_e)\tau_0\frac{e}{m_ec^2}v_{\nu}v^\mu F^{\alpha\nu}\right]\;,
\label{PLADdamped3}
\end{equation}
The strong damping parameter $\psi$  can be modified to include this quantum effect to obtain a  parameter $\psi_Q  = \langle g(\chi_e)\rangle\psi$, where $\langle g(\chi_e)\rangle$ is the time average of the $g$ factor. 
To do this, we make use of a  polynomial fraction fit to data for $g(\chi_e)$ given in Ref \cite{Kirk_PPCF_2009}:
\begin{eqnarray}
g(\chi_e) = \left(3.7\chi_e^3+31\chi_e^2+12\chi_e+1\right)^{-4/9}\;,
\label{gfactor}
\end{eqnarray}
which for $\chi_e\rightarrow0$, $g\rightarrow 1$. The condition $\chi_e =1$ corresponds to $g(\chi_e)=0.18$, but even for $\chi_e=0.1$ this factor has a value of $g(\chi_e)=0.66$. The time averaged field strength parameter $a_0/\sqrt{2}$ (for linear polarization) is used to approximate  $ \langle g(\chi_e)\rangle\approx g(\langle\chi_e\rangle)$, which is valid for $\chi_e\ll1$.

\begin{figure}[htbp]
\begin{center}
\includegraphics[width=0.5\textwidth]{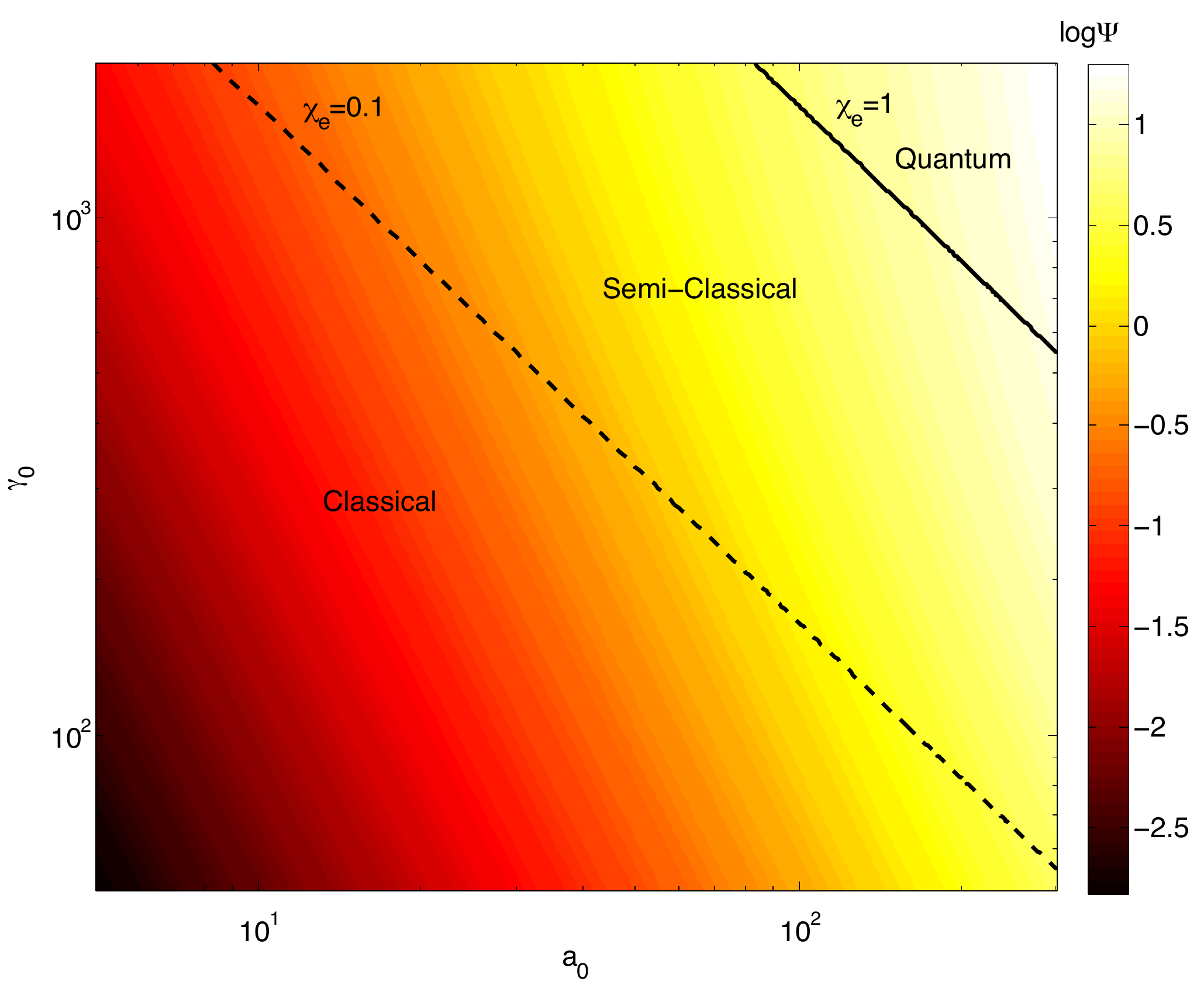}
\caption{The function $\psi_Q$ as a function of $a_0$ and $\gamma_0$ for an 800 nm central wavelength laser. The solid line indicates the threshold between  classical and quantum radiation reaction forces, and the dotted line indicates the threshold where $g(\chi_e)$ begins to be significant.}
\label{figure0}
\end{center}
\end{figure}

The modified strong damping parameter $\psi_Q$ for an 800~nm wavelength laser is shown as a function of $a_0$ and $\gamma_0$ in figure \ref{figure0}. As shown in Refs \cite{Kirk_PPCF_2009,Sokolov_PRE_2010}, for $\chi_e\sim0.1$, the spectrum emitted should not be changed significantly in shape, but $ \langle g(\chi_e)\rangle\approx g(\langle\chi_e\rangle)$ indicates that the energy loss of the electron beam due to radiation damping should be changed by a measurable amount. This is also consistent with what we observe with our model.

\subsection{Parameter regimes involving $\chi_e$ and $\psi_Q$}
The counter-propagating geometry laser-electron beam experiment is  an excellent testbed for studying quantum electrodynamics  and strong radiation damping effects. This is because of the ability to choose between strongly radiation damped behavior ($\psi_Q\gtrsim1$), or fields that are quantum electrodynamically strong  ($\chi_e\gtrsim1$), \emph{or} a situation where both $\psi_Q\gtrsim1$ and $\chi_e\gtrsim1$ simultaneously; conditions where even more exotic effects may occur. These are controlled  through  variation of the laser field strength, $a_0$, central frequency, $\omega_0$,  and the electron beam energy, $\gamma_0 m_ec^2$. We can compare the requirements for $a_0$, $\omega_0$ and $\gamma_0$ for the interaction to be in the \textit{strong radiation damping regime} or \textit{quantum electrodynamics dominant regime} relevant to experiments using 30 fs class lasers.

Table \ref{example_params} shows parameters for different scenarios for strongly radiation damped ($\psi_Q\gtrsim1$) and quantum electrodynamically strong ($\chi_e\gtrsim1$) physics in a non-linear Thomson/Compton scattering geometry for an 800 nm laser pulse with intensity $I_{L}$ colliding with an electron beam with energy $E_{b}$. (a) corresponds to the SLAC experiment \cite{Burke_PRL_1997}. (b) corresponds to near term experiments using intense 30 fs lasers such as {\sc Hercules}  \cite{Yanovsky:2008}  or Astra Gemini \cite{Hooker_JPIV_2006} and a laser wakefield generated electron beam.  (c) corresponds to an `ideal' experiment using two laser beam lines (as Astra Gemini has) with the current maximum experimentally demonstrated laser intensity \cite{Bahk_OL_2004} and laser-wakefield accelerated electron  beam energy \cite{Leemans_NP_2006}. The SLAC experiment is shown for comparative purposes only since  the quasi-static field approximation is not valid for this case \cite{Hu_PRL_2010}.

\begin{table}[htdp]
\caption{Different scenarios for strong radiation damping ($\psi_Q\gtrsim1$) and QED strong ($\chi_e\gtrsim1$) physics in a non-linear Thomson scattering geometry for 800 nm  wavelength laser pulses with intensity $I_{L}$ colliding with an electron beam with energy $E_{b}$.}
\begin{center}
\begin{tabular}{|c|c|c|c|c|c|c|c|c|c|}
\hline
& $E_{b}$ / GeV & $I_{L}$ / Wcm$^{-2}$& $a_0$ & $a_{rad}$& $a_Q$& $\chi_e$ &$\psi$ &$\psi_Q$\\
\hline
(a) &46.6 & $1\times10^{18}$&0.5 &2.5&1.2&0.43 & 0.045 & 0.018\\
(b) &0.2 & $5\times10^{21}$&50 &46&420&0.12 & 1.2 & 0.74\\
(c) &1 & $2\times10^{22}$&100&21&84&1.2 & 23 & 3.7\\
\hline
\end{tabular}
\end{center}
\label{example_params}
\end{table}%

In the first case (a) the laser vector-potential is of the order of both $a_{rad}$ and $a_Q$. 
Case (b) corresponds to a situation where there will be strong radiation damping but quantum effects will be weak, $a_Q>a_0>a_{rad}$. In case (c) the laser is sufficiently intense for both radiation damping and quantum recoil to be manifest, $a_0>a_Q>a_{rad}$.

 Since our model is  classical -- that is to say involving equations of motion only -- it is restricted to the parameter range where $\chi^2\ll 1$  \cite{Sokolov_PRE_2010}. For the parameters described here, $\chi^2=0.014$ and so the classical approach is reasonable. This also motivates the use of the description of this process as ``non-linear Thomson scattering'' rather than ``Compton'' scattering. We also calculate the electron spectrum after the interaction in the presence of radiation damping with and without the $g$ factor, showing that quantum modifications to radiation losses may be measurable.

\section{The model and numerical methods}
The spectral intensity of radiation emitted by a number $N_P$ of accelerating point charges can be expressed, in the far-field, as \cite{Jacksonbook}:
 \begin{equation}
\frac{d^2I}{d\omega d\Omega}=\frac{\mu_0e^2c}{16\pi^3}\omega^2\Bigg|{\int_{-\infty}^{\infty}\sum_{j=1}^{N_P}
\vect{\hat{s}}\times\vect{\beta}_je^{i\omega (t -\vect{n}\cdot \vect{r}_j/c)}}dt\Bigg|^2\;,
\label{spec:Jackson} 
\end{equation}
where the unit vector $\vect{\hat{s}}$ is in the direction of observation, at a distance far compared with the scale of the emission region. This can be written alternatively in terms of proper time, $\tau$:
 \begin{equation}
\frac{d^2I}{d\omega d\Omega}=\frac{\mu_0e^2c}{16\pi^3}\omega^2\Bigg|\sum_{j=1}^{N_P}{\int_{-\infty}^{\infty}\vect{\hat{s}}\times\vect{v}_je^{i\kappa_\alpha x_j^\alpha}}d\tau\Bigg|^2\;,
\label{spec:Jackson_proper} 
\end{equation}
where $\vect{v}_j$ is the momentum part of the $j$th particle's four-velocity defined as:
\begin{equation}
v_j^\alpha=\frac{dx_j^\alpha}{d\tau}\;.
\end{equation}

To numerically integrate the equations of motion  for charged particles, both $x^\alpha$ and $v^\alpha$ have to be recorded at a number of discrete points. To then perform the spectral integration numerically, equation \ref{spec:Jackson_proper} can be reduced to the summation:
 \begin{equation}
\frac{d^2I}{d\omega d\Omega}=\frac{\mu_0e^2c}{16\pi^3}\omega^2\Bigg|{\sum_{j=1}^{N_P}\sum_{n=0}^{N_\tau}\vect{\hat{s}}\times \vect{v}_j^ne^{i\kappa_\alpha x_j^{\alpha,n}}}\Delta\tau\Bigg|^2\;,
\label{spec:Jackson_proper_diff} 
\end{equation}
One of the advantages of using proper time rather than `laboratory' time for numerical calculations is that the time resolution is effectively adaptive; as the particle gains inertia and is therefore accelerated at a decreased rate for a similar force, the time step-size increases. Numerically calculating this integral by `brute force' has the problem that the exponent is a fast oscillating function, and therefore without resolving $\omega$ the integral will in general not converge without a numerical timestep of $\Delta \tau\ll1/\omega$ \cite{Filon_PRSE_1928,Nyquist,ISI:A1949um84500004}. Since we are interested in $\gamma$-ray photons in excess of an MeV energy generated from a few fs interaction, the ratio of the necessary time step to the integration timescale is computationally unfeasibly large. Recently, methods for overcoming this limit by using interpolation have been developed \cite{Martins_SPIE_2009,Thomas_PRSTAB_2010,Martins_AAC_2010}. Here we use the method we previously developed \cite{Thomas_PRSTAB_2010}, and the reader is directed towards that paper for further details of the numerical calculation.

The particle trajectories were calculated in the presence of four-potentials, $A^\mu = \{A^0=\phi/c,A^1,A^2,A^3\}$, representative of a spatio-temporally Gaussian laser pulse with no interaction between  electrons. The laser pulse propagated in the $+\vect{\hat{x}}_3$ direction with four-potential described by:
\begin{equation}
	A^\mu  =  \Re \left[\{A_0\}^\mu (x^\alpha) e^{i\kappa_{\alpha} x^\alpha}f(\kappa_{\alpha} x^\alpha/\omega_0t_L)\right]\;,
	\label{eq:laser_model}
\end{equation}
where $\{A_0\}^\mu (x^\alpha)$ is the spatial distribution of four-potential, in this case $\kappa_{\alpha}=\{\omega_0,0,0,\omega_0/c\}$ is the laser four wavevector, $f(\kappa_{\alpha} x^\alpha/\omega_0t_L)$ is a function of time describing the temporal envelope. The spatial-temporal distribution of a tightly focused pulse that satisfies the vacuum Maxwell's equations is in general very complicated, but is easier to formulate in terms of potentials than fields. That is because it is possible to have a purely transverse (to propagation) vector potential and satisfy the vacuum Maxwell's equations, something that is not possible with fields. These can be formulated by using the Lorentz-invariant Lorenz gauge condition $\partial_\mu A^\mu=0$. Using a slowly varying envelope approximation, and using a transverse vector potential linearly polarized in the $\hat{\vect{x}}_1$ direction with propagation in the $\hat{\vect{x}}_3$ direction, this can be approximated as $\{A_0\}^0= -(ic/\omega_0)\partial \{A_0\}^1/\partial x_1$  \cite{Davis_PRA_1979}. Here, vector and scalar potentials with corrections to the basic Gaussian optics formulation were introduced up to order ${\theta_0}^2$, where $\theta_0=2c/\omega_0w_0$ is the asymptotic divergence angle of a Gaussian laser beam with a waist of $w_0$. This yields potentials:

\begin{equation}
	\{A_0\}^1  =  \left[1+\frac{\theta_0^2}{2}\left(\frac{1-i\zeta}{1+\zeta^2}\right)\left(1-\left(\frac{1-\zeta^2}{2\left(1+\zeta^2\right)}\right)\rho^2\right)\right]\Psi_0\;,
	\label{eq:laser_model_A_1}
\end{equation}
\begin{equation}
	\{A_0\}^0  =  i\theta_0\xi_1e^{-i\tanh^{-1}\zeta}\left[\{A_0\}^1-{\theta_0}^2\left(\frac{1-\zeta^2}{1+\zeta^2}\right)\Psi_0\right]\;,
	\label{eq:laser_model_A_0}
\end{equation}
and $\{A_0\}^2=\{A_0\}^3=0$, where
\begin{equation}
	\Psi_0=\frac{e^{-i\tanh^{-1}\zeta-(1+i\zeta)\rho^2}}{\sqrt{1+\zeta^2}}\;,
	\label{eq:laser_model_Psi}
\end{equation}
$\rho=\sqrt{{x_1}^2+{x_2}^2}/w_0$, and $\zeta=x_3\theta_0/w_0$. $w_0/\theta_0$ is the Rayleigh range of the laser. Higher order corrections to the field structure could be employed to account for extremely tight focusing, but here we restrict our numerical calculations to foci with $w_0>\lambda$, where $\lambda$ is the laser wavelength; these corrections in ${\theta_0}^2$ are of magnitude $(1/\pi^2)\lambda^2/w_0^2$, so these corrections are up to 10~\% of the zero order fields and can't be considered negligible, but the next order corrections are ${\theta_0}^4$ and therefore of less importance.

An electron beam was modeled using $N_P$ particles initiated with a momentum $p_0$ in the $-\vect{\hat{x}}_3$ direction in front of the laser. In order to simulate a more realistic beam, rejection sampling against a Gaussian probability distribution function was used to generate a beam with a spread in momentum, $\vect{\sigma_p}$, and position, $\vect{\sigma_x}$ which statistically approximated the phase space distribution:
\begin{eqnarray}
&f_e(\vect{x},\vect{p},t) = &\exp\left[-\frac{x^2}{2\sigma_x^2}-\frac{p^2}{2\sigma_{p}^2}\right]\;,
\end{eqnarray}
where $x^2/\sigma_x^2 = x_1^2/\sigma_{x_1}^2+x_2^2/\sigma_{x_2}^2+x_3^2/\sigma_{x_3}^2$ and $p^2/\sigma_p^2 = p_1^2/\sigma_{p_1}^2+p_2^2/\sigma_{p_2}^2+\left(p_3-p_0\right)^2/\sigma_{p_3}^2$
The  root-mean-square normalized emittance of the bunch is therefore given by $\epsilon={\sigma_{p_1}}{\sigma_{p_2}}{\sigma_{p_3}}{\sigma_{x_1}}{\sigma_{x_2}}{\sigma_{x_3}}$. Although the particle tracking routine could easily calculate a much larger bunch, due to the computational demands of the numerical spectrometer for a full angular sweep, the number of electrons in the bunch was limited to $N_P=500$. Radiation from individual electrons was summed incoherently.

A  gaussian temporal envelope is used  in all cases, $f=e^{-(\kappa_{\alpha} x^\alpha/\omega_0t_L)^2}$. The pulse duration is $t_L=65\;\omega_0$, where $\omega_0$ is the laser angular frequency, which in the case of a typical $0.8$~$\mu$m laser is $2.36\times10^{15}$~s$^{-1}$, yielding $t_L=27.5$~fs at $1/e^2$ radius, or 32~fs full-width-at-half-maximum, of intensity. The electron beam parameters were varied, with $p_0$ corresponding to a beam energy typically of 204~MeV ($\gamma_0=400$). The linearly polarized laser, with  normalized vector potential of $a_0=50$ corresponding to a peak laser intensity of $5.3\times10^{21}$~Wcm$^{-2}$ was focused to a spot with waist $w_0=2.55$~$\mu$m, or $w_0=20c/\omega_0$. 
The electron beam parameters are comparable those routinely achieved in laser wakefield acceleration experiments and are summarized in table \ref{emit_tab}. 
\begin{table}[htdp]
\caption{Emittance parameters of the electron beam used in the numerical model. (a) Finite momentum spread case, (b) Zero momentum spread case.}
\begin{center}
\begin{tabular}{|c|c|c|c|c|c|c|}
\hline
&$\sigma_{x_1}$&$\sigma_{x_2}$&$\sigma_{x_3}$&$\sigma_{p_{1}}$&$\sigma_{p_{2}}$&$\sigma_{p_{3}}$\\
\hline
A&$3c/\omega_0$&$3c/\omega_0$&$9c/\omega_0$&$m_ec$&$m_ec$&$10m_ec$\\
B&$3c/\omega_0$&$3c/\omega_0$&$9c/\omega_0$&$0$&$0$&$0$\\
\hline
\end{tabular}
\end{center}
\label{emit_tab}
\end{table}%

\section{Numerical results}
In this section we detail `real world'  numerical calculations of a backscattering experiment applicable to near term experiments using current laser systems and laser wakefield accelerated electrons. By `real world', we mean that the calculation of the radiation spectrum includes the effect of a Gaussian shaped bunch of electrons with normalized emittance (longitudinal and transverse) comparable to that produced in laser-wakefield acceleration interacting with a tightly focused laser, that radiation reaction forces are included, and that the radiation spectrum is calculated directly from the electron trajectories. However, the self-consistent absorption of laser pulse photons is not included. The energy radiated by a $10^9$ electron beam will later be shown to be 0.3 J, which is a non-negligible 2\% of the pulse energy of the laser considered here. Including the depletion of laser energy would modify the spectrum of photons slightly, but is likely to be less important than the other effects we consider here.

\begin{figure}[htbp]
\begin{center}
\includegraphics[width=0.5\textwidth]{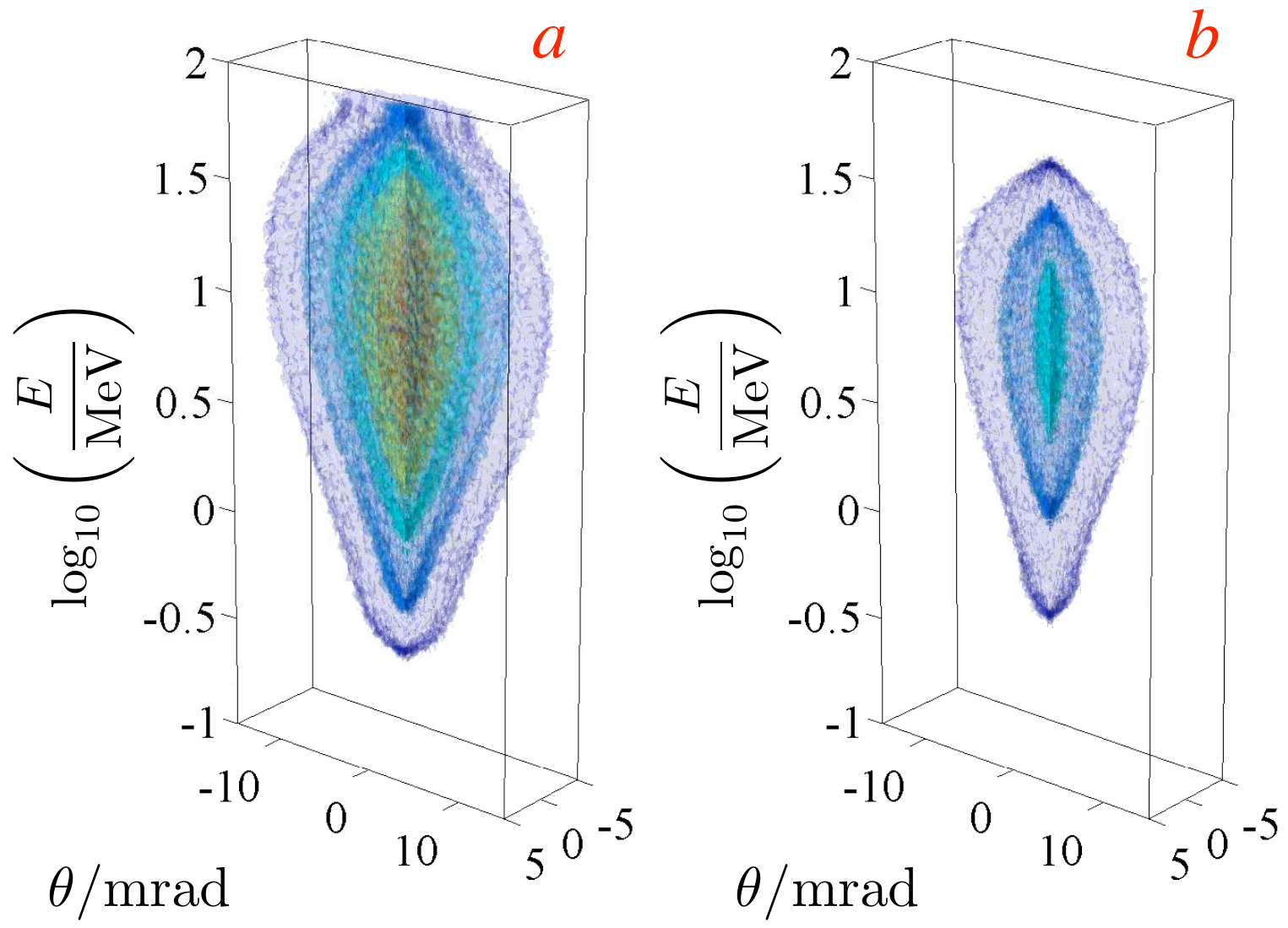}
\caption{The angularly resolved spectral intensity ($d^2I/d\omega d\Omega$) due to a 500 electron bunch with $\gamma=400$ and emittance given in table \ref{emit_tab} scattering from a laser pulse with $a_0=50$ {with higher order field contributions} included as in equations \ref{eq:laser_model_A_1}-\ref{eq:laser_model_Psi}. In (a) the radiation reaction force is not included, and in (b) the radiation reaction force is included. The contours are taken at identical spectral intensity levels for both cases, normalized to the peak which is 1.6517$\times10^{-26}$~Js$^{-1}$ at 0.2, 0.3, 0.4, 0.5, 0.6, 0.7 and 0.8.}
\label{figure1}
\end{center}
\end{figure}

The spectral intensity $d^2I/d\omega d\Omega$, where differential solid angle $d\Omega =\sin\theta d\theta d\phi$, was calculated on a grid consisting of 150 cells in $\omega$ over the range $10^4\omega_0<\omega<10^8\omega_0$, with $\Delta\omega$ exponentially increasing with $\omega$, 117 cells in $\theta$ over the range $0<\theta<30$ mrad and 26 cells in $\phi$ over the range $0<\phi<\pi/2$ rad. For clarity in the figures, symmetry is assumed and therefore the full range $-30<\theta<30$ mrad and $0<\phi<\pi$ is displayed.
\subsection{The high brilliance synchrotron source}
The properties of radiation from backscattering of an electron with a laser pulse have been extensively studied, and we can therefore use analytic formulae to predict that in the interaction of a electron beam with $\gamma=400$ with a laser with field strength $a_0=50$, the synchrotron-like spectrum will peak in energy at $\hbar\omega_{peak} = 2.56a_0\gamma^2\hbar\omega_0 = 30$~MeV \cite{ISI:000182450200081}. Using the numerical model, we can more accurately model the properties of the radiation produced in a high intensity laser interaction with a laser wakefield accelerated electron beam as a source for applications, in particular including the effects of radiation damping \cite{Koga_POP_2005}, non-plane wave laser fields \cite{Hartemann_AJS_2000} and calculate the full angular distribution of radiation. 

Figure \ref{figure1} shows the spectral intensity of radiation produced by a 500 electron bunch with $\gamma=400$  scattering from a laser pulse with $a_0=50$. In this example, the higher order field contributions are included, as in equations \ref{eq:laser_model_A_1}-\ref{eq:laser_model_Psi}, as well as beam emittance as given in table \ref{emit_tab} case A. This represents reasonably ``realistic'' modeling of an experiment and results in a well collimated, smooth, synchrotron-like radiation emission extending up to very high energies, with a broad peak at approximately $10$ MeV, which is a factor of 3 smaller than the analytic prediction due to the radiation reaction and finite spot effects. Because the laser pulse is linearly polarized, as expected, the radiation is strongly polarized, but also the angular intensity distribution has a pronounced ellipticity, with the major axis in the direction of polarization. Linear polarization also leads to higher photon energies compared with a circularly polarized pulse with the same pulse energy. 

One other notable effect is that of the higher order terms in the laser fields. These do not significantly change the spectral shape, but do change the magnitude non-negligibly. Without the field contributions, the peak spectral intensity is 1.57$\times10^{-26}$~Js$^{-1}$, but with them it is 1.65$\times10^{-26}$~Js$^{-1}$, which is a 5\% difference. Although the order $\theta^2$ pulse potential corrections have been simply added to the first order potential -- so that the energy in the corrected pulse is higher than the uncorrected -- because the additional potential are $\sim10$\% of the first order potential, adding the corrections only represents a $\sim$1\% increase in pulse energy. The slight increase in pulse energy alone is too small to account for the increased radiation output alone. Instead it is the additional longitudinal motion due to these potentials that increases the spectral output.

\begin{figure}[htbp]
\begin{center}
\includegraphics[width=0.5\textwidth]{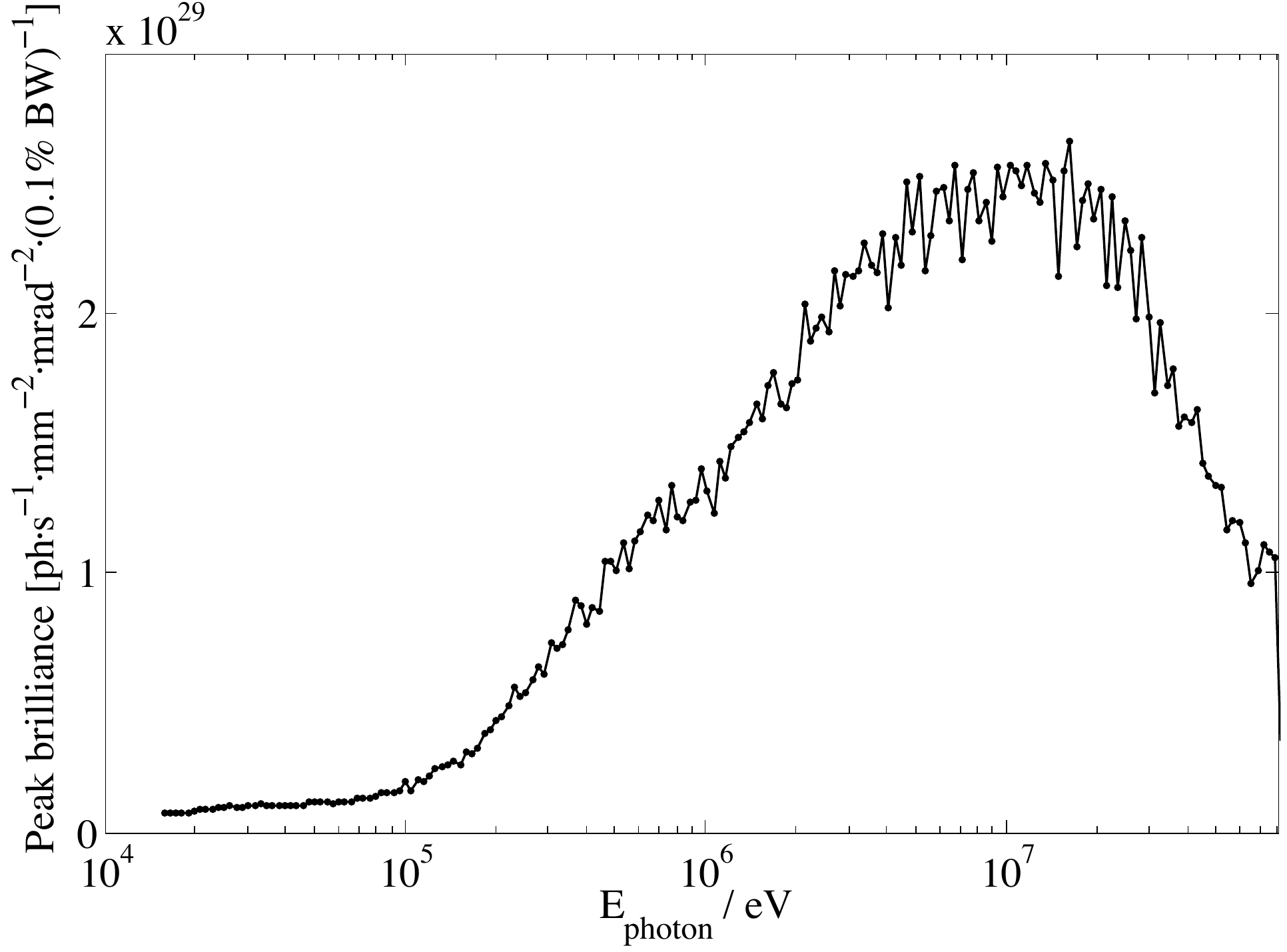}
\caption{The on-axis peak spectral brilliance ($ d^2I/d\omega d\Omega/(1000\hbar t_L\pi (w_0/2)^2)$ in standard light-source units of ${\rm photons}\cdot{\rm s}^{-1}{\rm mm}^{-2}{\rm mrad}^{-2} / 0.1\%\;{\rm bandwidth}$) due to a 100 pC electron bunch with $\gamma=400$ and momentum spread given by table \ref{emit_tab} scattering from a 30 fs laser pulse with $a_0=50$.}
\label{figure2}
\end{center}
\end{figure}

To compare this to other synchrotron light sources, it is also useful to plot the on-axis spectrum in terms of the standard units of the synchrotron community, ${\rm photons}\cdot{\rm s}^{-1}{\rm mm}^{-2}{\rm mrad}^{-2} / 0.1\%\;{\rm bandwidth}$. To this, the spectral intensity is multiplied by a numerical factor that assumes the 500 electrons are a reasonable statistical representation of a  100 pC electron bunch that is typical of laser wakefield experiments \cite{mangles,geddes,faure}. Also necessary for this calculation, the source size of the radiation is taken to be the laser spot area within the radius of half the pulse waist, $\pi (w_0/2)^2$. The on-axis radiation spectrum is shown in figure \ref{figure2}. As well as peaking at high energies, the peak spectral brilliance is also extremely high, comparable to the FLASH free electron laser but at significantly higher photon energies \cite{Robinson_NJP_2010} and significantly more brilliant than conventional synchrotrons. The effect of the high intensity dramatically increases the brilliance of the source, at the expense of the band-width which at lower intensity can be extremely narrow, which may be of more utility for some applications \cite{Hartemann_PRSTAB_2007,Albert_PRSTAB_2010,Albert_PRSTAB_2011}.

\begin{figure}[htbp]
\begin{center}
\includegraphics[width=0.5\textwidth]{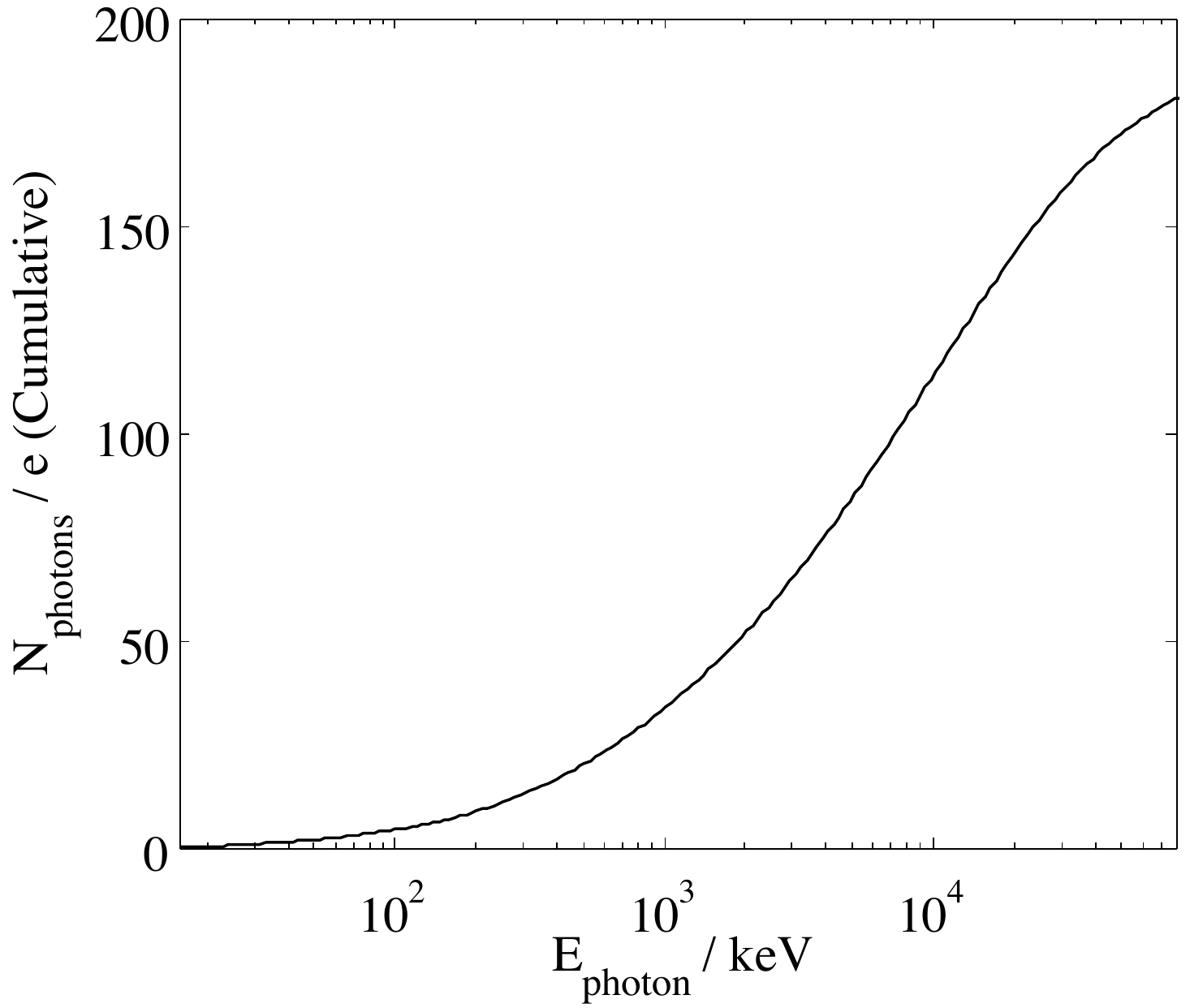}
\caption{The cumulative photon number ($\int_0^EdN/dE^\prime dE^\prime$) \textbf{per electron} due to a 500 electron bunch with $\gamma=400$ and momentum spread scattering given by table \ref{emit_tab} from a laser pulse with $a_0=50$.}
\label{figure3}
\end{center}
\end{figure}

 In figure \ref{figure3} the cumulative photon number, $\int_0^E(dN/dE^\prime) dE^\prime$, \emph{per electron} is shown for this spectrum, showing that on average each electron interacting with the laser field emits approximately 200 photons. When integrated numerically, the total photon energy emitted by each electron is $3.5\times10^{-10}$ J. This is 10 times more than the energy of a 200 MeV electron. This result may  superficially appear not to conserve energy, however, the radiated photon energy is predominantly drawn from the laser pulse.   For a bunch of $10^9$ electrons, which is of the order 100 pC of charge, the total energy output would be 0.35 J. For a bunch of this size, depletion of the laser fields -- if treated self-consistently -- would modify the electron dynamics and radiation output, but should only be  a small perturbation (the pulse energy used here is 19.0 J) and hence would not be expected to modify this output energy significantly. Ignoring this correction, the conversion efficiency of laser pulse energy into $\gamma$-rays is 1.8\%.

\begin{figure}[htbp]
\begin{center}
\includegraphics[width=0.5\textwidth]{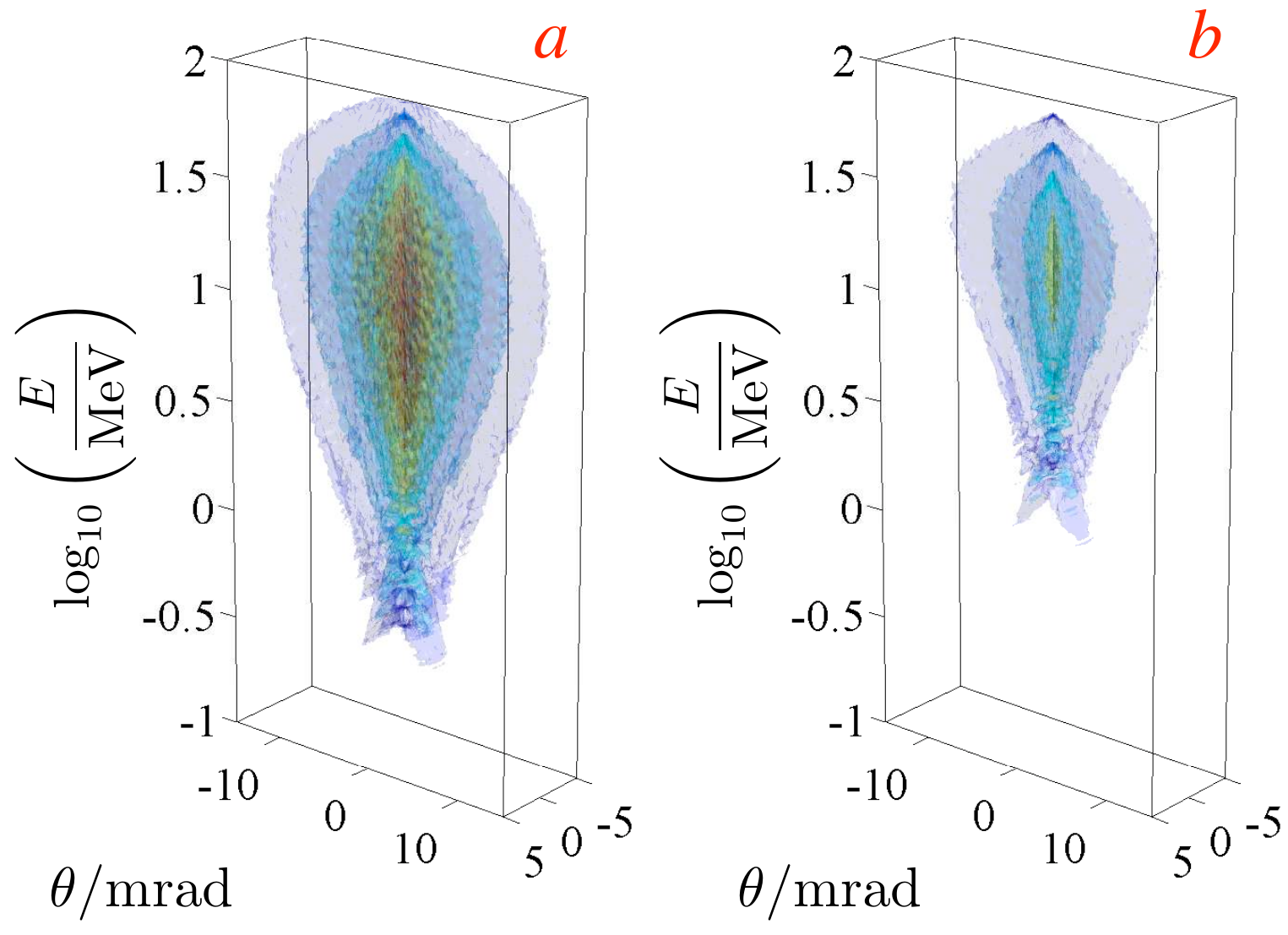}
\caption{The angularly resolved spectral intensity ($d^2I/d\omega d\Omega$) due to a zero emittance 500 electron bunch with $\gamma=400$ and no scattering from a laser pulse with $a_0=50$. In (a) the radiation reaction force is not included, and in (b) the radiation reaction force is included. The contours are taken at identical spectral intensity levels for both cases, normalized to the peak which is 2.6872$\times10^{-26}$~Js$^{-1}$ at 0.2, 0.3, 0.4, 0.5, 0.6, 0.7 and 0.8.}
\label{figure4}
\end{center}
\end{figure}

\subsection{On the observation of radiation reaction effects in the photon distribution}
\begin{figure}[htbp]
\begin{center}
\includegraphics[width=0.5\textwidth]{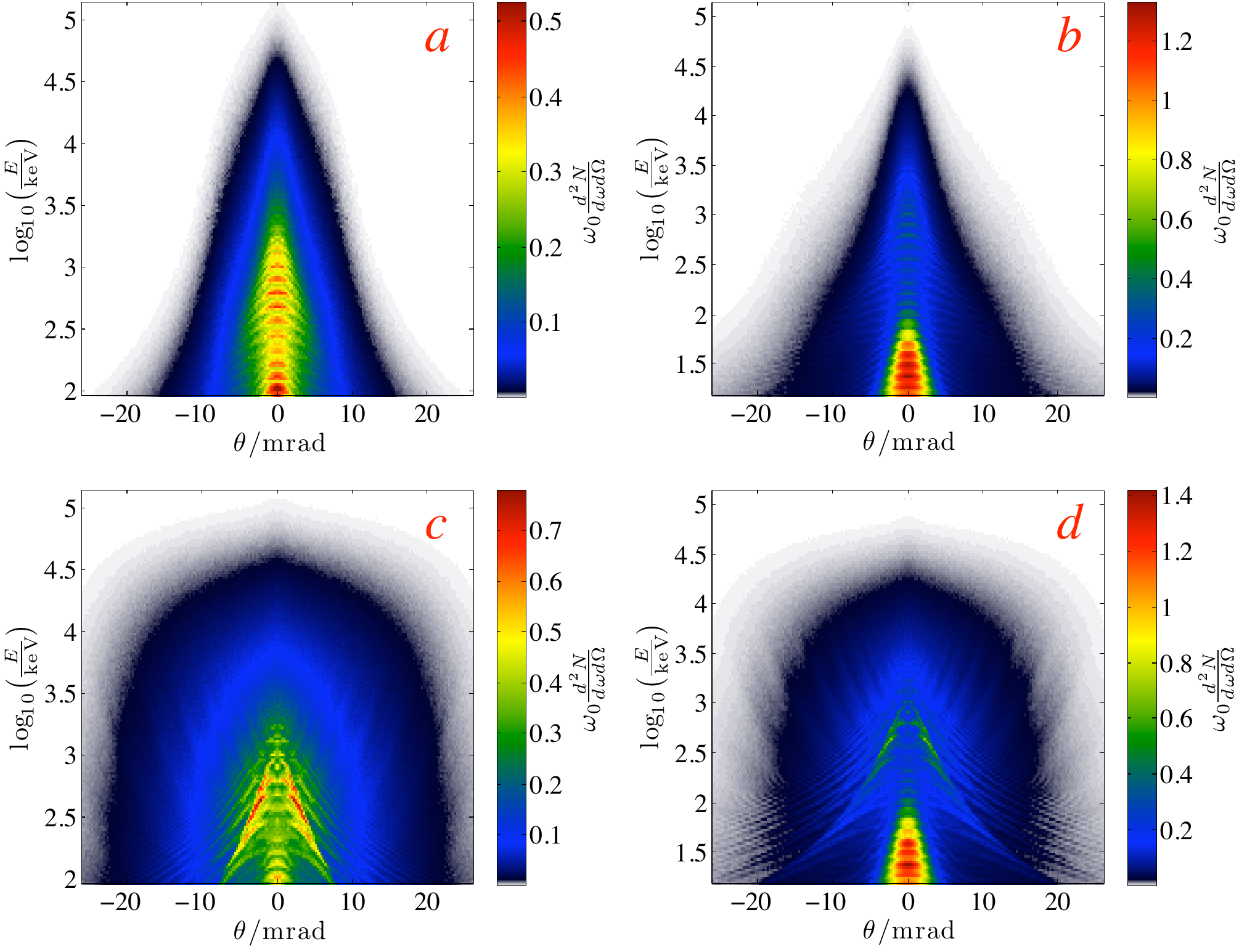}
\caption{The  photon distribution (normalized to the laser frequency, $\omega_0d^2N/d\omega d\Omega$)  \textbf{per electron} due to a 500 electron bunch with $\gamma=400$ and zero momentum spread scattering from a laser pulse with $a_0=50$. In (a) and (c) \textbf{radiation reaction force is not included}, and in (b) and (d) \textbf{radiation reaction force is included}. (a) and (b) show the photon distribution in the plane \textbf{perpendicular} to the laser polarization and (c) and (d) show the photon distribution in the plane \textbf{parallel} to the laser polarization.}
\label{figure5}
\end{center}
\end{figure}

\begin{figure}[htbp]
\begin{center}
\includegraphics[width=0.5\textwidth]{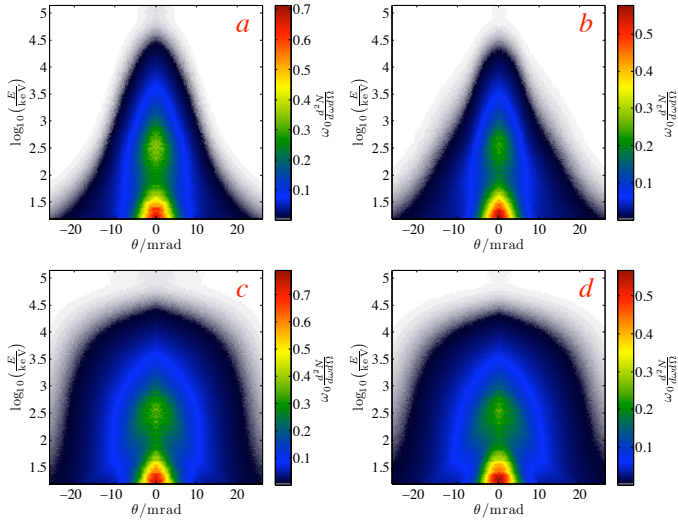}
\caption{The  photon distribution (normalized to the laser frequency, $\omega_0d^2N/d\omega d\Omega$) \textbf{per electron} due to a 500 electron bunch with $\gamma=400$ and momentum spread given by table \ref{emit_tab} scattering from a laser pulse with $a_0=50$. In (a) and (c) \textbf{radiation reaction force is not included}, and in (b) and (d) \textbf{radiation reaction force is included}. (a) and (b) show the photon distribution in the plane \textbf{perpendicular} to the laser polarization and (c) and (d) show the photon distribution in the plane \textbf{parallel} to the laser polarization.}
\label{figure6}
\end{center}
\end{figure}

It has been suggested  that signatures of the radiation reaction forces may be observed in the photon distribution emitted in a counter propagating experiment \cite{Koga_POP_2005,Piazza_PRL_2009}. The numerical calculations performed here suggest this may be difficult due to the momentum spread of the electron beam. Figure \ref{figure4} shows the spectral intensity of radiation emitted under identical conditions to those of figure \ref{figure1} except that here the electron beam has zero momentum spread,  as in table \ref{emit_tab} case B. The distribution has fine features that are smoothed out when the electron beam has a momentum spread, as would be expected. To see more clearly the effect of momentum spread on the radiation distribution, figures \ref{figure5} and \ref{figure6} show two dimensional slices through the radiation intensity distribution, in the planes parallel and perpendicular to the laser polarization. In addition, the spectral intensity has been converted into a photon distribution \emph{per electron}, $\omega_0d^2N/d\omega d\Omega$, which is more likely to be the form of data obtained in an experiment (i.e. a histogram of photon hits on an array of single photon counting detectors).

 Figure \ref{figure5} shows the photon distribution from a zero momentum spread electron beam interaction. (a) and (c) \emph{radiation reaction force is not included}, and in (b) and (d) \emph{radiation reaction force is included}. (a) and (b) show the photon distribution in the plane {perpendicular} to the laser polarization and (c) and (d) show the photon distribution in the plane {parallel} to the laser polarization. The angular distribution of photons shows pronounced differences with and without radiation reaction forces, and the energy distribution is also dramatically changed, in particular resulting in a large number of low energy photons in the damped case compared to no damping. Another feature is slow oscillations in the spectral intensity with frequency/energy. These may be due to the short truncated electron bunch and laser pulse in the time domain, which will result in long wavelength oscillations in the frequency domain. 

When the electron bunch is given the momentum spread of table \ref{emit_tab} case A, the distinction between the cases with and without radiation reaction force becomes significantly less distinct. Figure \ref{figure6} shows the photon distribution from this interaction. There is  little difference in the spectral intensity distribution with and without radiation reaction force effects, except that the overall magnitude is reduced, and the peak energy is reduced. Differences in the angular distribution, however, are small and are likely to be much smaller than expected shot-to-shot fluctuations in electron beam emittance. Coupling this to the intrinsic difficulty of measuring high energy photons in a collimated beam, it appears to be unfeasible to suggest that radiation reaction effects will be discernible in experimental measurements in this configuration in the near term.

\subsection{On the observation of radiation reaction effects in the electron phase-space distribution}
In contrast to the photon measurements, it should be very easy to observe radiation reaction effects in the electrons as measured using a standard scintillating screen configuration. It is typical in laser wakefield accelerator experiments to measure either the electron beam profile using a scintillating screen, or electron forward momentum spectrum using a deflecting magnet and a scintillating screen \cite{geddes,faure}. These diagnostics effectively correspond to the $p_1$-$p_2$ and $p_1$-$p_3$ electron phase-space densities respectively --  with a spectrometer, the deflection by the magnetic field disperses the electrons by $p_3$, but the projection in $p_1$ is maintained.

\begin{figure}[htbp]
\begin{center}
\includegraphics[width=0.5\textwidth]{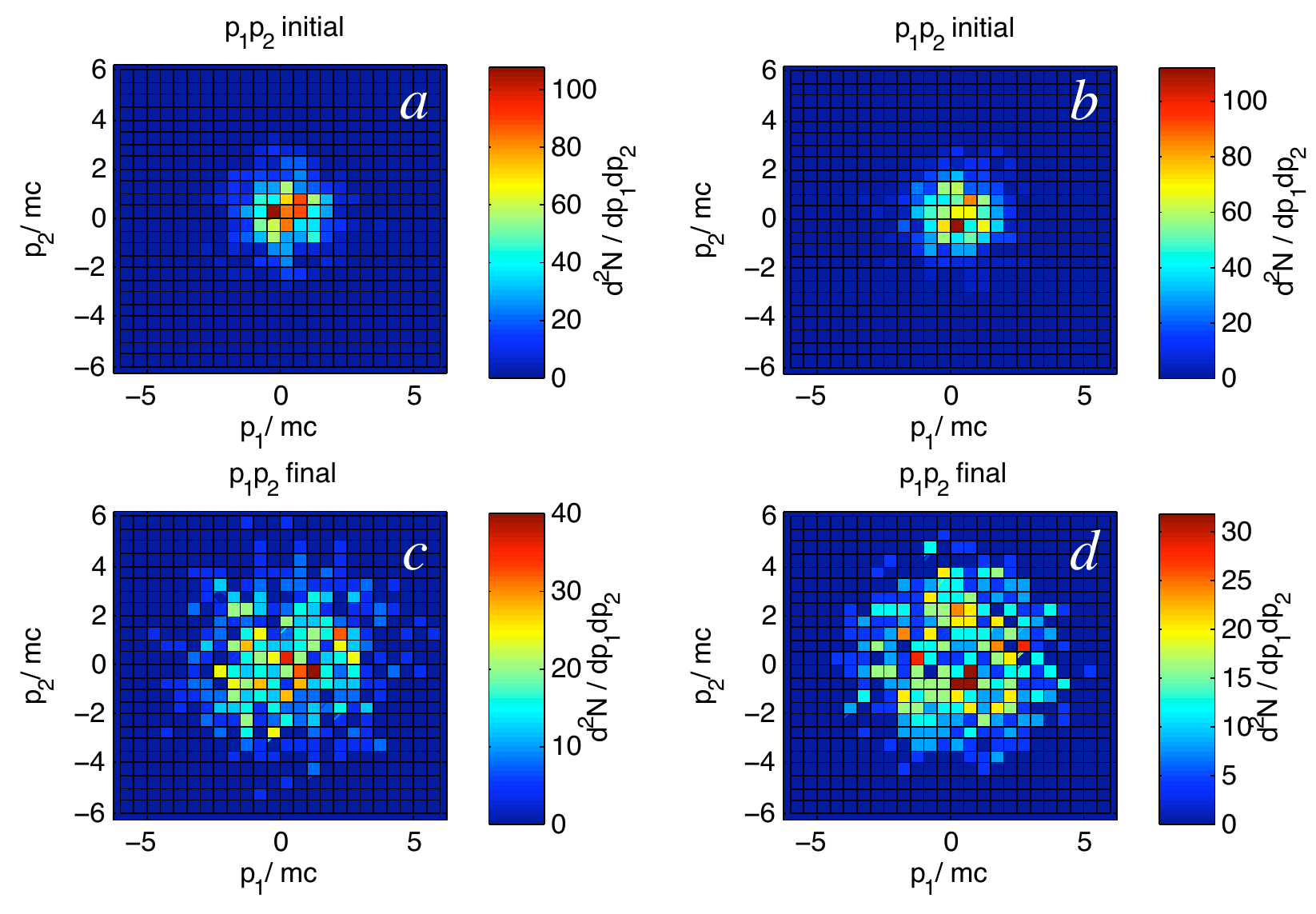}
\caption{Two dimensional histograms of the $p_1$-$p_2$  phase space distribution of a 500 electron bunch with {momentum spread according to table \ref{emit_tab}} case B before (top) and after (bottom) interaction with the high intensity ($a_0=50$) pulse. In (a) and (c) there is no radiation reaction, and in (b) and (d) a radiation reaction model included according to equation \ref{PLAD}.}
\label{figure7}
\end{center}
\end{figure}

Figure \ref{figure7} shows the  $p_1$-$p_2$ phase-space density for the electron bunch before and after the interaction as two dimensional histogram plots. The electrons are deflected by the laser fields so that the transverse momentum spread is increased in both cases, consistent with a ponderomotive deflection. However, there is little difference between the cases with and without radiation reaction forces. This is because the radiation damping effect reduces both transverse and longitudinal momenta proportionally (to lowest order the radiation force in  equation \ref{PLAD} is $dp^\mu/d\tau|_{fric} = -\tau_0\omega_0^2\gamma^2a^2p^\mu$), and hence in general the exit angle of a particular electron $\theta_{exit}\simeq p_\bot/p_3$ is not expected to  change significantly. 

\begin{figure}[htbp]
\begin{center}
\includegraphics[width=0.5\textwidth]{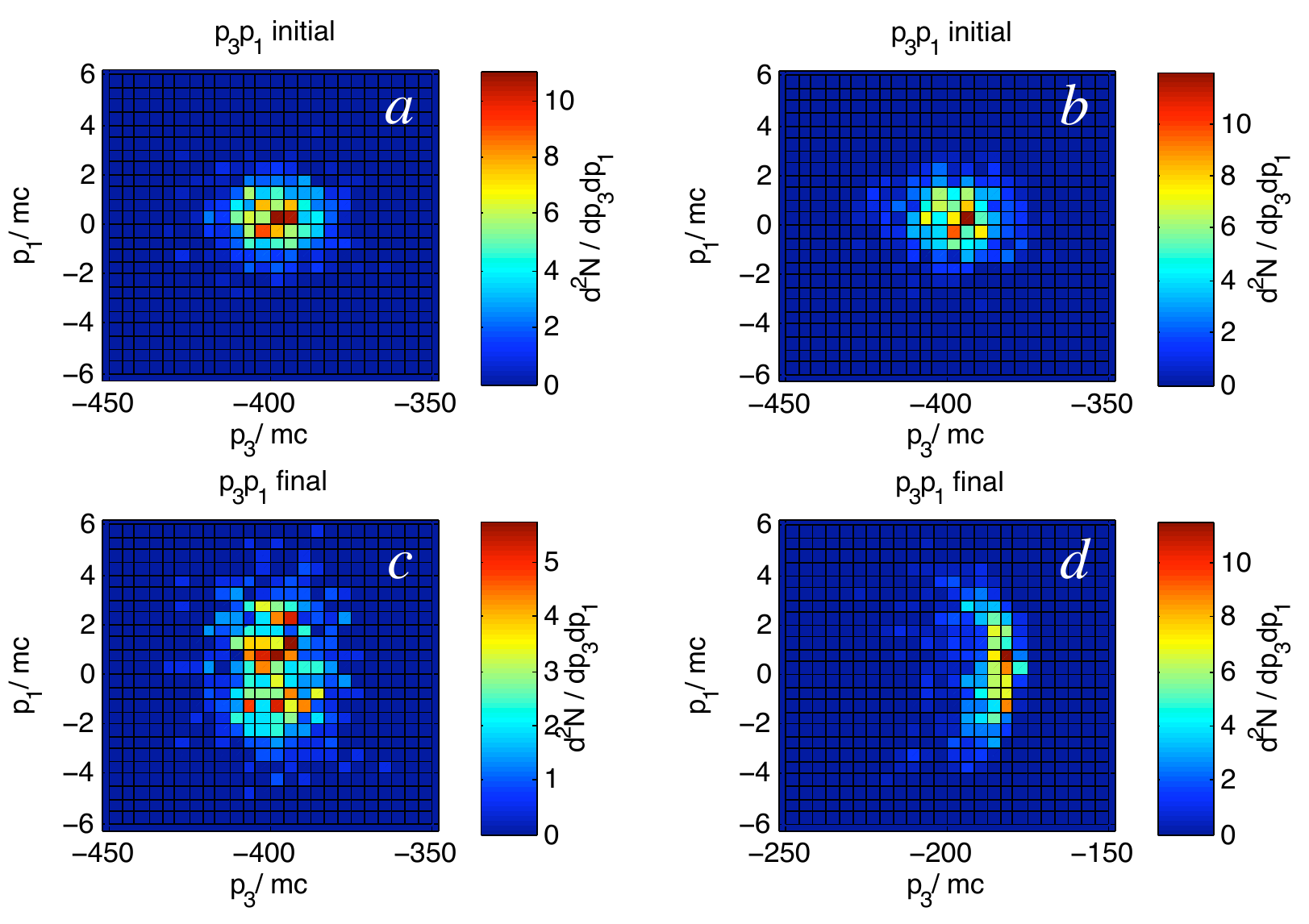}
\caption{Two dimensional histograms of the $p_3$-$p_1$  phase space distribution of a 500 electron bunch with {momentum spread according to table \ref{emit_tab}} case B before (top) and after (bottom) interaction with the high intensity ($a_0=50$) pulse. In (a) and (c) there is no radiation reaction, and in (b) and (d) a radiation reaction model included according to equation \ref{PLAD}. \emph{Note that the horizontal momentum scale is negative and not the same for each phase space}.}
\label{figure8}
\end{center}
\end{figure}

The effect on the electron spectrum is dramatic however, as was also previously shown by Koga et al. \cite{Koga_POP_2005}. In figure \ref{figure8}, two dimensional histogram plots of the $p_3$-$p_1$ phase space density of the electron bunch is shown with and without radiation reaction forces.  Under the conditions modeled here, the electron beam loses almost half its energy when radiation damping is included (and as expected, it experiences little change in energy without radiation damping).

\begin{figure}[htbp]
\begin{center}
\includegraphics[width=0.5\textwidth]{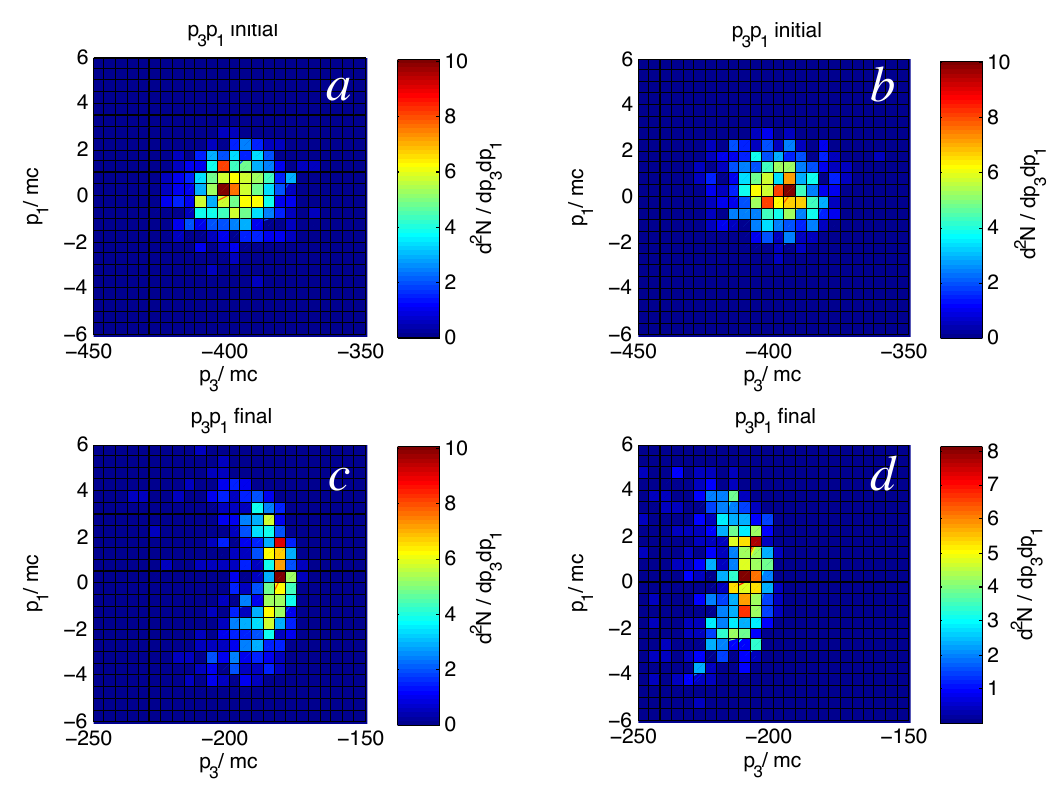}
\caption{Two dimensional histograms of the $p_3$-$p_1$  phase space distribution of a 500 electron bunch with {momentum spread according to table \ref{emit_tab}} case B before (top) and after (bottom) interaction with the high intensity ($a_0=50$) pulse with radiation damping.  (a) and (c) use the purely classical expression \ref{PLAD} and in (b) and (d) the radiation reaction four-force is modified by multiplying by the instantaneous $g$ factor given by equation \ref{gfactor}.  \emph{Note that the horizontal momentum scale is negative and not the same for each phase space.}}
\label{figure9}
\end{center}
\end{figure}

Finally, figure \ref{figure9} shows  two dimensional histogram plots of the $p_3$-$p_1$ phase space density of the electron bunch  similar to figure \ref{figure8}, but this time the radiation reaction is shown with and without the factor   $g(\chi_e)$ given by equation \ref{gfactor} included. It can be seen by the plot that under these conditions the \emph{electron} spectrum after the interaction  with the $g$ factor differs from the purely classical result by $\sim10$\% relative to the overall energy loss. There is a smaller energy loss due to the fact that the expected radiation spectrum is less energetic than the purely classical result would suggest. This difference between classical and quantum corrected radiation reaction may be sufficiently large to be distinguishable over experimental fluctuations if well characterized. The effect of the addition of $g(\chi_e)$ on the \emph{photon} spectrum calculated  with classical radiation reaction forces under these conditions is negligible. 

\section{Conclusions}
The counter-propagating electron beam, ultra-high-intensity laser interaction is likely to be attempted by numerous experimental groups in the near future. In addition to ultimately studying quantum electrodynamic effects, initial experiments with lower electron beam energies and laser intensities are likely to be concerned with the brilliant high energy photon output and classical forms of radiation forces. From these numerical calculations, we predict a large flux of photons with energy in excess of 1 MeV, in a beam collimated within a 10~mrad divergence angle, and with an elliptical angular distribution due to the linear polarization of the laser pulse. Each electron should emit $\sim$100 photons above 1 MeV for a 200 MeV Gaussian electron beam colliding with a pulse of intensity $5\times10^{21}$~Wcm$^{-2}$. For a typical laser wakefield accelerated electron bunch with 100 pC charge \cite{mangles,geddes,faure}, this should result in $\sim10^{11}$ photons in a broad synchrotron-like spectrum peaking at 10 MeV with approximately 2\% conversion efficiency of laser energy into $\gamma$-rays, in a beam collimated to less than 10 mrad divergence and with a peak brightness exceeding $10^{29}$ photons$\,$s$^{-1}$mm$^{-2}$mrad$^{-2}(0.1$\% bandwidth$)^{-1}$.

In addition we show that measurements of the \emph{radiation} will be unlikely to be able to indicate signatures of radiation reaction forces, and in particular the ability to distinguish between different classical or quantum formulations of the radiation force, due to the effects of beam emittance and tight laser focusing. However, it should still be easy to observe radiation reaction effects in the electron spectrum, where differences compared with a no-radiation force model is dramatic, even with moderate beam emittance. Including quantum effects using the $g(\chi_e)$ factor under these parameters causes a sufficiently reduced damping effect on the electron energy spectrum to be measurable. 

Whether signatures of different classical radiation reaction force models can be observed in experiment is not addressed by the results of this paper. However, it is unlikely, since the primary measurable effect on the electrons is energy loss, which is to low order similar for all formulations of the radiation reaction force. It is likely that the differences between models will be hidden by the effects of beam emittance and laser focusing conditions also.
\section{Acknowledgements}
 This work was funded by the NSF under contracts 1054164 and PHY-0935197, DARPA under contract N66001-11-1-4208, EPSRC (Grant numbers EP/I014462/1 and EP/G055165/1) and the Royal Society.


\end{document}